\pdfoutput=1
\documentclass[aip,pop,amsmath,amssymb,reprint,groupedaddress]{revtex4-1}

\usepackage{graphicx} 
\usepackage{dcolumn}
\usepackage{bm}
\usepackage{amssymb}
\usepackage{amsmath}
\usepackage{wasysym}
\usepackage{color}
\usepackage{hyperref}	
\usepackage[toc,page]{appendix}
\usepackage{float}
\usepackage{flushend}
\usepackage{hyperref}
\usepackage{tabularx}
\hypersetup{
    colorlinks,%
    citecolor=blue,%
    filecolor=blue,%
    linkcolor=blue,%
    urlcolor=blue
} 

\definecolor{Gray}{gray}{0.90}
\newcolumntype{a}{>{\columncolor{Gray}}c}

\begin{document}
             
\title{Discrete Symmetries in Dynamo Reversals}
\author{Riddhi Bandyopadhyay}
\thanks{Present address: Department of Physics and Astronomy, University of Delaware, Newark, DE 19716, USA}

\author{Mahendra K. Verma}
\email{mkv@iitk.ac.in}
\affiliation{Department of Physics, Indian Institute of Technology Kanpur- 208016, India}
\date{\today}
             
\begin{abstract}
Quantification of the velocity and magnetic field reversals in dynamo remains an interesting challenge.  In this paper, using group-theoretic analysis,  we  classify the reversing and non-reversing Fourier modes during a dynamo reversal in a Cartesian box.  Based on odd-even parities of the wavenumber indices, we categorise the velocity and magnetic Fourier modes into 8 classes each.  Then, using the properties of the nonlinear interactions in magnetohydrodynamics, we show that these 16 elements form Klein 16-group $Z_2 \times Z_2 \times Z_2 \times Z_2$.   We demonstrate that field reversals in a class of Taylor-Green dynamo, as well as the reversals in earlier experiments and models, belong to one of the classes predicted by our group-theoretic arguments. 

\end{abstract}
\maketitle

\section{Introduction}

In 1919, Larmor proposed that the magnetic field in various astrophysical and geophysical bodies are generated self-inductively by the electric currents and magnetic field by a bootstrap mechanism~\cite{Moffat:book}.  This mechanism is called {\em dynamo}.  The generated magnetic field exhibits many interesting phenomena including field reversals.   Paleomagnetic records show that the Earth's magnetic field has  reversed its polarity on geological time scales~\cite{brunhes:jpa1906}.  The interval between two reversals is random with an average interval between two consecutive reversals as approximately 200,000 years. On the contrary, the magnetic field of the Sun changes its polarity  quasi-periodically approximately every eleven years~\cite{Moffat:book}.  This phenomena called {\em field reversal} is an interesting puzzle, and it has been studied by a large number of researchers.  In this paper we will study the symmetry properties of such reversals.

Various theoretical models  have been proposed to describe dynamo mechanism suitable for different situations. Magnetohydrodynamics (MHD), which treats the plasma as fluid, is often used to describe the behaviour of turbulent plasma in the presence of magnetic field. The MHD model however breaks down for collisionless and relativistic limits, and other models are employed for such cases.  Dynamo actions in collisionless plasmas have been recently investigated by Rincon {\em et al.}~\cite{Rincon:PNAS2016} and Kunz {\em et al.}~\cite{Kunz:PRL2016}. This is pertinent to dynamos in extragalactic plasmas, e.g., in accretion disks, intercluster medium etc.~\cite{Zeldovich:book:Magneticfield} In addition, Hall effect becomes important when the ion and electron velocities are sufficiently distinct.  Mininni {\em et al.}~\cite{Mininni:ApJ2002,Mininni:ApJ2003a,Mininni:ApJ2003b,Mininni:ApJ2005,Mininni:JPP2007}, G\'omez {\em et al.}~\cite{Gomez:PRE2010}, and Lingam and Bhattacharjee~\cite{Lingam:ApJ2016} have investigated Hall dynamos.  There are other possibilities of dynamo action, but we do not list them here due to limited scope of this paper.   In this paper, for simplicity, we limit ourselves to MHD dynamos.

The equations of magnetohydrodynamics (MHD) satisfy the symmetry properties: ${\bf u \rightarrow u}$ and  ${\bf b \rightarrow -b}$, where ${\bf u,b}$ are the velocity and magnetic fields respectively.    Note however that such symmetry is not persevered in generalised MHD, such as Hall MHD.  From the above symmetry of MHD, one may infer that the magnetic field  changes sign after a reversal in MHD, but the velocity field does not.  However, researchers observe that   ${\bf b \rightarrow -b}$ in some experiment, but in some others, only some of the large-scale modes of the ${\bf b}$ switch sign, and some others do not. In the laboratory experiment involving a Von Karman swirling flow of liquid Sodium (VKS)~\cite{Berhanu:EPL2007}, the magnetic dipolar component $D$ reverses  but the magnetic quadrupolar component $Q$ does not.  Petrelis {\em et al.}~\cite{Petrelis:JOP_Cond_Mat2008} and Gissinger {\em et al.}~\cite{Gissinger:PRE2010} constructed low-dimensional models whose variables are   dipolar and quadrupolar magnetic fields, and  dipolar velocity field.  Gissinger~\cite{Gissinger:PRE2010} showed that the field reversals in VKS experiment is consistent with the predictions of his low-dimensional models.

 Earlier, Rikitake~\cite{Rikitake:MathProc1958} , Nozieres~\cite{Nozieres:PEARI1978} , and Knobloch~\cite{Knobloch:PLA1981} studied magnetic field reversals in disk dynamo model~\cite{Moffat:book} and its variations.  These models also exhibit chaos.  Tobias {\em et al.}~\cite{Tobias:mnras1995} studied chaotically-moduled stellar dynamo. Refer to Moffatt~\cite{Moffat:book}, Weiss and Proctor~\cite{weiss_proctor:book} and reference therein for further discussions on low-dimensional dynamo models that exhibit   reversals of magnetic field. 

Rayleigh-B\'{e}nard convection (RBC) exhibits flow reversals in which the velocity field reverses randomly in time~\cite{Niemela:JFM2001,Ahlers:JFM2006,Xi:PRE2006}. The dynamics of flow reversal has significant similarities with that of dynamo reversals.   For the flow reversal in a two-dimensional (2D) box, Chandra and Verma~\cite{Chandra:PRE2011, Chandra:PRL2013} and Verma {\em et al.}~\cite{Verma:POF2015} constructed group-theoretic arguments to identify the  Fourier modes that change sign in 2D flow reversals of RBC.  They showed that the reversing and non-reversing Fourier modes of 2D RBC form a Klein four-group $Z_2 \times Z_2$.  In Appendix A, we generalise the above arguments to three-dimensional (3D) RBC.  In this paper we make similar symmetry-based arguments for the dynamo reversals, and classify Fourier modes that change sign in a dynamo reversal.  The group consists of 8 elements each of the velocity and magnetic fields.  We discuss the details of the group structure in Sec.~\ref{sec:Sym}.  

It is important to contrast the reversal dynamics  observed in box, cylinder, or sphere geometries.  For RBC in a box, flow reversals are accompanied by sign changes of some of the Fourier modes~\cite{Sugiyama:PRL2010, Chandra:PRE2011, Chandra:PRL2013}.  In a cylindrical convection, similar  reversals have been observed, and they are referred to as {\em cessation-led reversal}.   In addition, cylindrical convection exhibits   {\em rotation-led reversal} in which the  vertical velocity field changes sign due to the rotation of the large-scale circulation~\cite{Brown:JFM2006, Xi:PRE2007, Xi:PRE2008, Mishra:JFM2011}.  In terms of symmetries, the reversals of flow structures in Cartesian geometry or those in a cessation-led reversal are related to the {\em discrete symmetry}.  But reversals due to continuous rotation, as in rotation-led reversals in a cylinder, are connected to the {\em continuous symmetry}.   It is natural to expect that both types of reversals would occur in a spherical dynamo. In the present paper we focus on discrete symmetries of dynamo.  Cylindrical and spherical geometries exhibit reversals  connected to discrete and continuous symmetries.  In this paper, to keep the focus on discrete symmetries,  we focus on dynamo reversals in a box geometry.

Some of the observational works related to dynamo reversals are discussed below. Earth's magnetic field, which is generated by the motion of molten iron inside the Earth,  has a dominant dipolar structure. Most of the Earth's past magnetic field data have been measured from the ferromagnet rocks that were formed out of the  frozen magma. Using these measurements, geologists discovered that the Earth's magnetic field has reversed many times in the past.   The interval between two consecutive field reversals is randomly distributed, and the  field structure during a reversal is quite complex with possible  multipolar magnetic-field structure.    Some scientists believe that the geomagnetic reversals   is a spontaneous process, while others argue it to be triggered by some external sources~\cite{Muller:GRL1986, Muller:GRL2002}.

The solar dynamo too exhibits polarity reversals, but these reversals differ significantly from the geomagnetic reversals.  The sunspots, solar wind, and solar flares provide us valuable inputs about the Sun's magnetic field.  For example, the poloidal field reverses its direction approximately every eleven years; a field reversal involves interactions among the   poloidal and toroidal components~\cite{Parker:Book}.  

Fast supercomputers  and sophisticated numerical codes have enabled researchers to simulate and study aforementioned dynamo mechanism in  realistic geometries, e.g. in spherical shells.  However the parameters used in simulations are quite far from the realistic values. Field reversals have been reported in several numerical simulations of the geodynamo~\cite{Roberts:RMP2000} and other 3D simulations of rotating spheres~\cite{Dormy:GGG2000}.  Glatzmaier and Roberts~\cite{Glatzmaier:NATURE1995} ran  simulations equivalent to approximately 300000 terrestrial years, and observed field reversals similar to those observed in paleomagnetic records.  They reported that the interval distribution between two consecutive reversals is random, and that the magnetic field geometry has a complex structure during a reversal.

Based on symmetry arguments, P\'{e}trelis {\em et al.}~\cite{Petrelis:PRL2009} proposed a mechanism for dynamo reversals in VKS.  They assumed that the magnetic field is decomposed into two parts---a dipolar component of amplitude $D$, and a quadrupolar component of amplitude $Q$, and constructed a variable $A$: 
\begin{eqnarray}
A = D + i Q. \label{eq:qd_decomp}
\end{eqnarray}
They wrote the following amplitude equation in powers of $A$ and its complex conjugate $\bar{A}$  under  the constraint that $\mathbf{B} \rightarrow - \mathbf{B}$ (or $A \rightarrow - A$):
\begin{eqnarray}
\dot{A} = \mu A + \nu \bar{A} + \beta_1 A^3 + \beta_2 A^2 \bar{A} + \beta_3 A \bar{A}^2 + \beta_4 \bar{A}^3. \label{eq:amp_exp}
\end{eqnarray}
This is up to the lowest order nonlinearity. Here, $\mu$, $\nu$, $\beta_i$s are complex coefficients that depend on the experimental parameters. Using this model, P\'{e}trelis {\em et al.}~\cite{Petrelis:PRL2009} explained various dynamic regimes of the VKS experiment.  

In a related development, Gissinger {\em et al.}~\cite{Gissinger:EPL2010} considered  a three-mode model of dynamo reversal.   A third mode $V$ representing the large-scale velocity is considered in addition to $D$ and $Q$.   The governing equations are derived based on symmetries, and they are
\begin{eqnarray}
\dot{D} = \mu D - V Q, \label{eq:D_eqn} \\
\dot{Q} = -\nu Q + V D,\\
\dot{V} = \Gamma - V - Q D \label{eq:qvd_eqn}
\end{eqnarray}
upto quadratic nonlinearities. A nonzero $\Gamma$ represents the forcing that breaks the rotational symmetry.  The models of P\'{e}trelis {\em et al.}~\cite{Petrelis:PRL2009} and Gissinger {\em et al.}~\cite{Gissinger:EPL2010} invoke rotation and mirror symmetries to construct the nonlinear terms and determine the reversing and non-reversing modes.  

In this paper we present group-theoretic arguments to determine the reversing and non-reversing modes in a dynamo reversal.  Our analysis exploit the nonlinear structure of the equation.  We will show that ${\bf u} \rightarrow {\bf u}$ and ${\bf b} \rightarrow -{\bf b}$ is a subclass of the possible reversals.   Our arguments show that some ${\bf u,b}$ modes reverse and some others do not.  Though our symmetry arguments are similar to those of  P\'{e}trelis {\em et al.}~\cite{Petrelis:PRL2009} and  Gissinger {\em et al.}~\cite{Gissinger:EPL2010}, yet they are more convenient due to their algebraic structure. Our model also encompasses more modes in contrast to a smaller number  of large-scale modes in the models of Gissinger {\em et al.}~\cite{Gissinger:EPL2010}.

Krstulovic \textit{et al.}~\cite{Krstulovic:PRE2011} simulated  Taylor-Green dynamo for various boundary conditions. They observed that the dynamo thresholds varies with the boundary conditions. In a box geometry with insulating walls, Krstulovic \textit{et al.}~\cite{Krstulovic:PRE2011} observed  an axial dipolar dynamo similar to that  in VKS experiment~\cite{Monchaux:PRL2007}.  However Krstulovic \textit{et al.}~\cite{Krstulovic:PRE2011} did not study the dynamo reversals.  In this paper we extend Krstulovic \textit{et al.}'s study so as to include dynamo reversals.  We observed interesting reversals for the insulating boundary condition; here the dipolar mode does not flip, but higher Fourier modes flip.  We will show that the set of reversing and non-reversing modes belong to one of the solutions of the group-theoretic model.

Kutzner and Christensen~\cite{Kutzner:PEP2002} performed direct numerical simulations (DNS) of MHD equations, and observed transitions between dipolar and multipolar regime accompanied by reversals of the dipolar field. Oruba and Dormy~\cite{Oruba:GRL2014} showed that such transitions from the static dipolar to the reversing multipolar dynamo are due to balance between the inertial, viscous and Coriolis forces.  These investigations raise interesting question on the reversing and non-reversing modes in a dynamo.    The general symmetry classes of our group-theoretic arguments would be useful for such analysis.

The outline of the paper is as follows: in Sec.~\ref{sec:GE}, we discuss the  governing equations and the boundary conditions of the system. In Sec.~\ref{sec:Sym}, we describe the symmetries of dynamo reversals.  We extend these arguments to magneto-convection in Sec.~\ref{sec:magneto_convection}. In Sec.~\ref{sec:DNS} we show that the reversing and non-reversing modes in a dynamo reversals observed in a DNS belong to one of the classes of group-theoretic model. In Sec.~\ref{sec:earlier_dynamos} we discuss the symmetry classes of some other dynamo reversals.  We conclude in the Sec.~\ref{sec:Con}.

\section{Equations and Method}\label{sec:GE}  
The governing equations of a dynamo are
\begin{eqnarray} 
\frac{\partial \mathbf{u}}{\partial t} + (\mathbf{u} \cdot \nabla)\mathbf{u} & = &  - \nabla P + (\mathbf{j} \times \mathbf{b}) + \nu \nabla^2 \mathbf{u} + \mathbf{f}, \label{eq:u_non}\\ 
\frac{\partial \mathbf{b}}{\partial t} + (\mathbf{u} \cdot \nabla)\mathbf{b} & = & (\mathbf{b} \cdot \nabla)\mathbf{u} + \eta \nabla^2 \mathbf{b}, \label{eq:induc}\\
\nabla \cdot \mathbf{u} & = & 0, \label{eq:inc}\\
\nabla \cdot \mathbf{b} & = & 0 \label{eq:b_gauss}.
\end{eqnarray}
where $\mathbf{u}$ is the velocity field, $\mathbf{b}$ is the magnetic field, $P$ is the pressure field, $\mathbf{f}$ is the external mechanical forcing, $\mathbf{j}=(\nabla \times \mathbf{b})/\mu_0$ is the current density, $\nu$ is the kinematic viscosity, and $\eta$ is the magnetic diffusivity. We consider the flow to be incompressible [see Eq. (\ref{eq:inc})], and set the fluid density $\rho$ to unity.

Two important parameters used in a dynamo literature are the Reynolds number $\mathrm{Re} $ and magnetic Reynolds number $\mathrm{Rm}$, which are defined as
\begin{eqnarray}
\mathrm{Re} & = & \frac{U  L}{\nu},\\
\mathrm{Rm} & = & \frac{U L}{\eta},
\end{eqnarray}
where $U = \sqrt{2 E_u}$ is the root-mean-square velocity ($E_u$ = the total kinetic energy), and $L$ is the characteristic length scale of the flow, which is defined as
\begin{align}
L &= \frac{2\pi \int k^{-1} E_u(k) dk}{\int E_u(k) dk}.
\end{align}
Here the one-dimensional kinetic energy spectrum $E_u(k)$ is defined as the energy contents of a shell of radius $k$ and unit width:
\begin{equation}
E_u(k) = \sum_{k-1 < |\mathbf{k}^\prime| \le k}\frac{1}{2} |\hat{\mathbf{u}}(\mathbf{k}^\prime)|^2.
\end{equation}
The one-dimensional magnetic energy spectrum is defined similarly.
 One other important dimensionless parameter for dynamo is  magnetic Prandlt number, which is defined as
\begin{equation}
\mathrm{Pm} = \frac{\nu}{\eta} = \frac{\mathrm{Re}}{\mathrm{Rm}}.
\end{equation}

For the analysis of  the large-scale structures and dynamo reversals, it is convenient to work in the Fourier space with Fourier basis function $\exp(i{\bf k} \cdot {\bf r})$:
\begin{eqnarray}
{\bf u} & = & \hspace*{-0.3cm}\sum_{k_x,k_y,k_z}    \hat{{\bf u}} (k_x,k_y,k_z) \exp(i {\bf k \cdot x})   \label{eq:uk_fourier}, \\
{\bf b} & = & \hspace*{-0.3cm}\sum_{k_x,k_y,k_z}    \hat{{\bf b}} (k_x,k_y,k_z) \exp(i {\bf k \cdot x}),   \label{eq:bk_fourier}
\end{eqnarray}
where $k_x,k_y,k_z$ are integers for a $(2\pi)^3$ box; they take both positive and negative values.  In this representation,  the MHD equations are
\begin{align}
\frac{d}{dt} \hat{u}_m(\mathbf{k}) & =  -i k_n \hspace*{-0.1cm} \sum_{\mathbf{k}=\mathbf{p}+\mathbf{q}} \hspace*{-0.2cm} \hat{u}_n(\mathbf{q}) \hat{u}_m(\mathbf{p}) + i k_n \hspace*{-0.1cm} \sum_{\mathbf{k}=\mathbf{p}+\mathbf{q}} \hspace*{-0.2cm} \hat{b}_n(\mathbf{q}) \hat{b}_m(\mathbf{p}) \nonumber\\ &- \nu k^2 \hat{u}_m(\mathbf{k}) - i k_m \hat{p}(\mathbf{k}) + \hat{f}_m(k)\label{eq:u_k},
\end{align}
\begin{align}
\frac{d}{dt}  \hat{b}_m(\mathbf{k}) & =  -i k_n \hspace*{-0.1cm} \sum_{\mathbf{k}=\mathbf{p}+\mathbf{q}} \hspace*{-0.2cm} \hat{b}_n(\mathbf{q}) \hat{u}_m(\mathbf{p}) + i k_n \hspace*{-0.1cm} \sum_{\mathbf{k}=\mathbf{p}+\mathbf{q}} \hspace*{-0.2cm} \hat{u}_n(\mathbf{q}) \hat{b}_m(\mathbf{p}) \nonumber\\ &- \eta k^2 \hat{b}_m(\mathbf{k})\label{eq:b_k},
\end{align}
\begin{align}
k_{m}\hat{u}_m(\mathbf{k}) & =  0\label{eq:inc_k} , \\
k_{m}\hat{b}_m(\mathbf{k}) & =  0\label{eq:b_gauss_k}. 				
\end{align}
where $\hat{u}_m(\mathbf{k})$, $\hat{b}_m(\mathbf{k})$, $\hat{f}(\mathbf{k})$, and $\hat{p}(\mathbf{k})$ are the Fourier transforms of the velocity, magnetic, external force, and pressure fields respectively.   We employ Taylor-Green (TG) forcing: 
\begin{align}
\mathbf{f} &= \text{F}_0  \begin{bmatrix}
 & \sin({k_0 x})\cos({k_0 y})\cos({k_0 z})    \\
 & - \cos({k_0 x})\sin({k_0 y}) \cos({k_0 z})   \\
 & 0   \end{bmatrix}  
 \label{eq:force}
\end{align}
where $\text{F}_0$ is the forcing amplitude, and $k_0$ is the wavenumber of that forcing.  We choose $k_0=1$.  

For the velocity field we employ the free-slip or stress-free boundary condition at all the six sides of the  box:
\begin{align}
\begin{split}
u_\perp = 0;\quad \frac{\partial u_\parallel}{\partial n}=0,
\end{split}
\end{align} 
where $\hat{n}$ is the normal to the surface, and ${\color{blue}{u_\perp}}$ and $u_\parallel$ are respectively the velocity components normal and parallel to the wall. For example, if we consider the boundary condition at the wall at $z=0$ ($xy$ plane), the normal vector $\hat{n}$ will be the $-\hat{z}$ vector. So, in this case, $u_\perp = -u_z = 0$,  $\partial u_x/\partial z = 0$, and $\partial u_y/\partial z = 0$. For the magnetic field, we employ the insulating boundary condition at all the walls~\cite{Krstulovic:PRE2011}:
\begin{align}
\begin{split}
b_\parallel = 0;\quad  \frac{\partial b_\perp}{\partial n}=0,
\end{split}
\end{align} 
where $b_\perp$ and $b_\parallel$ are respectively the components of the magnetic field, normal and parallel to the wall. We call this insulating wall   because the current  $\mathbf{j} = (\nabla \times \mathbf{b})/\mu_0$ on the surface is zero.  The aforementioned boundary conditions are satisfied for  the following basis functions for ${\bf u}$ and ${\bf b}$:
\begin{align}
u_{x} & = & \hspace*{-0.3cm}\sum_{k_x,k_y,k_z}\hspace*{-0.3cm}  8 \hat{\hat{u}}_{x}(k_x,k_y,k_z) \sin(k_{x}x)   \cos(k_{y}y) \cos(k_{z}z)    \label{eq:ux_3d}, \\
u_{y} & = & \hspace*{-0.3cm}\sum_{k_x,k_y,k_z}\hspace*{-0.3cm}  8 \hat{\hat{u}}_{y}(k_x,k_y,k_z) \cos(k_{x}x)   \sin(k_{y}y) \cos(k_{z}z)    \label{eq:uy_3d}, \\
u_{z} & = & \hspace*{-0.3cm}\sum_{k_x,k_y,k_z}\hspace*{-0.3cm}  8 \hat{\hat{u}}_{z}(k_x,k_y,k_z) \cos(k_{x}x) \cos(k_{y}y)  \sin(k_{z}z) \label{eq:uz_3d}, \\
b_{x} & = & \hspace*{-0.3cm}\sum_{k_x,k_y,k_z}\hspace*{-0.3cm}  8 \hat{\hat{u}}_{x}(k_x,k_y,k_z) \cos(k_{x}x)   \sin(k_{y}y) \sin(k_{z}z)    \label{eq:bx_3d}, \\
b_{y} & = & \hspace*{-0.3cm}\sum_{k_x,k_y,k_z}\hspace*{-0.3cm}  8 \hat{\hat{b}}_{y}(k_x,k_y,k_z) \sin(k_{x}x)   \cos(k_{y}y) \sin(k_{z}z)    \label{eq:by_3d}, \\
b_{z} & = & \hspace*{-0.3cm}\sum_{k_x,k_y,k_z}\hspace*{-0.3cm}  8 \hat{\hat{b}}_{z}(k_x,k_y,k_z) \sin(k_{x}x) \sin(k_{y}y)  \cos(k_{z}z) \label{eq:bz_3d},
\end{align}
  where, in a $\pi^3$ box,  $k_x, k_y, k_z$ are positive integers including zero, and $\hat{\hat{u}}, \hat{\hat{b}}$ represent the basis functions for the free-slip and insulating boundary conditions.   We choose the above basis functions for our simulation.    We refer to the above as {\em free-slip, insulating basis function}, for which we follow the conventions and definitions of FFTW~\cite{Frigo:IEEE2005}.  Using $2 i \sin({\bf k \cdot r}) = \exp({\bf k \cdot r}) -  \exp(-{\bf k \cdot r})$ and $2\cos({\bf k \cdot r} =\exp({\bf k \cdot r}) -  \exp(-{\bf k \cdot r})$, we can relate $ \hat{\hat{u}}_{i}(k_x,k_y,k_z) $ and $ \hat{\hat{b}}_{i}(k_x,k_y,k_z) $ with $\hat{{\bf u}} (\pm k_x, \pm k_y,\pm k_z)$ and $\hat{{\bf b}} (\pm k_x, \pm k_y,\pm k_z)$ of Eqs.~(\ref{eq:uk_fourier},\ref{eq:bk_fourier}). For example, $\hat{u}_{x}(-k_x,-k_y,-k_z) = - i \hat{\hat{u}}_{x}(k_x,k_y,k_z)$, $\hat{u}_{x}(-k_x,k_y,-k_z) =  \hat{\hat{u}}_{x}(k_x,k_y,k_z)$, etc.   These properties enable us to use 1/8th Fourier modes (of Eqs.~(\ref{eq:uk_fourier},\ref{eq:bk_fourier})) for a pseudo-spectral simulation. However, in our simulation, we impose the above condition in each time step, and time-step all the Fourier modes, i.e., $\hat{{\bf u}} (\pm k_x, \pm k_y,\pm k_z)$ and $\hat{{\bf b}} (\pm k_x, \pm k_y,\pm k_z)$ of Eqs.~(\ref{eq:uk_fourier},\ref{eq:bk_fourier}).

  The Fourier decomposition of the MHD equations yield a set of coupled nonlinear ordinary differential equations (ODEs) given by Eqs.~(\ref{eq:u_k},\ref{eq:b_k}). These equations are often solved numerically using pseudo-spectral method, as done in this paper (see Sec.~V).  It is also customary to truncate the Fourier expansion drastically and focus only on a limited set of modes. Hence we obtain a small set of  nonlinear ODEs that can be analysed using the tools of nonlinear dynamics.  The dimension of the system, and consequently, its complexity will depend on the order of truncation. Quantities like Lyapunov exponents can be used to study properties of such systems, e.g., the transition between deterministic and chaotic behaviour. Following the above procedure, Verma {\em et al.}~\cite{Verma:PRE2008} constructed a truncated six-model using the Fouier modes ${\bf u}(1,0,1), {\bf u}(0,1,1), {\bf u}(1,1,2), {\bf b}(1,0,1), {\bf b}(0,1,1), {\bf b}(1,1,2)$. These Fourier modes are part of an interacting triad.  The above model exhibits dynamo transition, but no chaos.

In the next section, we discuss the symmetries of the MHD flows; these symmetries provide valuable insights  into the dynamo reversals.

\section{Symmetries of the MHD equations and participating modes}
\label{sec:Sym}

The structure of the MHD equations~\cite{Moffat:book} reveal that the equations are invariant under the transformation ${\bf u \rightarrow u}$ and  ${\bf b \rightarrow -b}$.  However, in several dynamo simulations and models, only some modes of the velocity and magnetic fields reverse during a dynamo reversal, e.g., in Gissinger~\cite{Gissinger:PRE2010}, the dipolar component of the magnetic field reverses, but not the quadrupolar component. The rules for such reversals can be derived using the symmetry properties of the MHD equations in the Fourier space.  The arguments are somewhat simpler for the velocity field only, which appears in RBC.  In Appendix A, we discuss the symmetry properties for RBC. 
 
For the basis functions of Eqs.~(\ref{eq:ux_3d}-\ref{eq:bz_3d}), we classify the Fourier modes according to the parity of the modes (${\bf k} = (k_x, k_y, k_z)$).  For the same, we divide each Fourier component, $k_x, k_y, k_z$, according to their parities---even, represented by $e$, and odd, represented by $o$.   Let us denote the parity function by $P$. To illustrate, $P(3) = o$, but $P(4) = e$.  Thus, for MHD, the Fourier modes of the velocity field (under the free-slip boundary condition) is classified into eight classes: $E= (eee)$, odd $O=(ooo)$, and mixed modes---$M_1=  (eoo)$, $M_2 =  (oeo)$, $M_3=  (ooe)$, $M_4=  (eeo)$, $M_5=  (oee)$, $M_6= (eoe)$. The corresponding classes for the magnetic Fourier modes are labeled as $\overline{E}$, $\overline{O}$, $\overline{M}_1$, $\overline{M}_2$, $\overline{M}_3$, $\overline{M}_4$, $\overline{M}_5$, and $\overline{M}_6$ respectively.  To illustrate, $\hat{u}_x(1,1,1) \in O$ , $\hat{u}_x(2,2,2) \in E$, and $\hat{b}_x(2,1,1) \in \overline{M}_1$. In the following discussion, we will derive a group structure for the above modes that will help us identify the reversing and nonreversing Fourier modes.

 \begin{table*}
\caption{Rules of nonlinear interactions among the Fourier modes of MHD. The elements form an abelian Klein-16 group $Z_2 \times Z_2 \times Z_2 \times Z_2$.} 
\label{tab:3d_product_rule}
  \renewcommand{\arraystretch}{1.5}
  \renewcommand{\tabcolsep}{8.0pt}
  \noindent
\begin{tabularx}{\textwidth}{  >{$}l<{$} | >{$}l<{$}  >{$}l<{$}  >{$}l<{$}  >{$}l<{$}  >{$}l<{$}  >{$}l<{$}  >{$}l<{$}  >{$}l<{$} |  >{$}l<{$}  >{$}l<{$}  >{$}l<{$}  >{$}l<{$}  >{$}l<{$}  >{$}l<{$}  >{$}l<{$}  >{$}l<{$}  >{$}l<{$} } 
\hline \hline
\times & E & O & M_{1} & M_{2} & M_{3} & M_{4} & M_{5} & M_{6}  & \overline{E} & \overline{O} & \overline{M}_{1} & \overline{M}_{2} & \overline{M}_{3} & \overline{M}_{4} & \overline{M}_{5} & \overline{M}_{6} \\ 
  \hline 
 E & E & O & M_{1} & M_{2} & M_{3} & M_{4} & M_{5} & M_{6} & \overline{E} & \overline{O} & \overline{M}_{1} & \overline{M}_{2} & \overline{M}_{3} & \overline{M}_{4} & \overline{M}_{5} & \overline{M}_{6}  \\
 
 O & O & E & M_{5} & M_{6} & M_{4} & M_{3} & M_{1} & M_{2} & \overline{O} & \overline{E} & \overline{M}_{5} & \overline{M}_{6} & \overline{M}_{4} & \overline{M}_{3} & \overline{M}_{1} & \overline{M}_{2}  \\
 
 M_{1} & M_{1} & M_{5} & E & M_{3} & M_{2} & M_{6} & O & M_{4} & \overline{M}_{1} & \overline{M}_{5} & \overline{E} & \overline{M}_{3} & \overline{M}_{2} & \overline{M}_{6} & \overline{O} & \overline{M}_{4}  \\
 
 M_{2} & M_{2} & M_{6} & M_{3} & E & M_{1} & M_{5} & M_{4} & O & \overline{M}_{2} & \overline{M}_{6} & \overline{M}_{3} & \overline{E} & \overline{M}_{1} & \overline{M}_{5} & \overline{M}_{4} & \overline{O}  \\
 
 M_{3} & M_{3} & M_{4} & M_{2} & M_{1} & E & O & M_{6} & M_{5} & \overline{M}_{3} & \overline{M}_{4} & \overline{M}_{2} & \overline{M}_{1} & \overline{E} & \overline{O} & \overline{M}_{6} & \overline{M}_{5}  \\
 
 M_{4} & M_{4} & M_{3} & M_{6} & M_{5} & O & E & M_{2} & M_{1} & \overline{M}_{4} & \overline{M}_{3} & \overline{M}_{6} & \overline{M}_{5} & \overline{O} & \overline{E} & \overline{M}_{2} & \overline{M}_{1} \\
 
 M_{5} & M_{5} & M_{1} & O & M_{4} & M_{6} & M_{2} & E & M_{3} & \overline{M}_{5} & \overline{M}_{1} & \overline{O} & \overline{M}_{4} & \overline{M}_{6} & \overline{M}_{2} & \overline{E} & \overline{M}_{3} \\
 
 M_{6} & M_{6} & M_{2} & M_{4} & O & M_{5} & M_{1} & M_{3} & E & \overline{M}_{6} & \overline{M}_{2} & \overline{M}_{4} & \overline{O} & \overline{M}_{5} & \overline{M}_{1} & \overline{M}_{3} & \overline{E} \\   \hline 
 
\overline{E} & \overline{E} & \overline{O} & \overline{M}_{1} & \overline{M}_{2} & \overline{M}_{3} & \overline{M}_{4} & \overline{M}_{5} & \overline{M}_{6} & E & O & M_{1} & M_{2} & M_{3} & M_{4} & M_{5} & M_{6}  \\

\overline{O} & \overline{O} & \overline{E} & \overline{M}_{5} & \overline{M}_{6} & \overline{M}_{4} & \overline{M}_{3} & \overline{M}_{1} & \overline{M}_{2} & O & E & M_{5} & M_{6} & M_{4} & M_{3} & M_{1} & M_{2}  \\

\overline{M}_{1} & \overline{M}_{1} & \overline{M}_{5} & \overline{E} & \overline{M}_{3} & \overline{M}_{2} & \overline{M}_{6} & \overline{O} & M_{4} & M_{1} & M_{5} & E & M_{3} & M_{2} & M_{6} & O & M_{4}  \\

\overline{M}_{2} & \overline{M}_{2} & \overline{M}_{6} & \overline{M}_{3} & \overline{E} & \overline{M}_{1} & \overline{M}_{5} & \overline{M}_{4} & \overline{O} & M_{2} & M_{6} & M_{3} & E & M_{1} & M_{5} & M_{4} & O  \\

\overline{M}_{3} & \overline{M}_{3} & \overline{M}_{4} & \overline{M}_{2} & \overline{M}_{1} & \overline{E} & \overline{O} & \overline{M}_{6} & \overline{M}_{5} & M_{3} & M_{4} & M_{2} & M_{1} & E & O & M_{6} & M_{5}  \\

\overline{M}_{4} & \overline{M}_{4} & \overline{M}_{3} & \overline{M}_{6} & \overline{M}_{5} & \overline{O} & \overline{E} & \overline{M}_{2} & \overline{M}_{1} & M_{4} & M_{3} & M_{6} & M_{5} & O & E & M_{2} & M_{1} \\

\overline{M}_{5} & \overline{M}_{5} & \overline{M}_{1} & \overline{O} & \overline{M}_{4} & \overline{M}_{6} & \overline{M}_{2} & \overline{E} & \overline{M}_{3} & M_{5} & M_{1} & O & M_{4} & M_{6} & M_{2} & E & M_{3} \\

\overline{M}_{6} & \overline{M}_{6} & \overline{M}_{2} & \overline{M}_{4} & \overline{O} & \overline{M}_{5} & \overline{M}_{1} & \overline{M}_{3} & \overline{E} & M_{6} & M_{2} & M_{4} & O & M_{5} & M_{1} & M_{3} & E \\
\hline \hline
\end{tabularx}
\end{table*}

\begin{table}[htp]
\caption{For dynamo, the classes of Reversing modes (R) and Non-reversing  (NR) Fourier modes.  The remaining modes have small amplitudes and they fluctuate around zero. }
\label{tab:MHD_rules}
\begin{center}
\begin{tabular}{c|c|c|c}
\hline
Item & R  & NR  & No of classes  \\ \hline 
1 & one $b$ element   & $E$   & ${8}\choose{1}$ = 8 classes \\
2 & two $b$ elements & $E$    & ${8}\choose{2}$ =28  classes \\
3 & three $b$ elements & $E$   &  ${8}\choose{3}$ =56  classes \\
4 & four $b$ elements & $E$    & ${8}\choose{4}$ =28  classes \\
5 & five $b$ elements & $E$    & ${8}\choose{5}$ =56  classes \\
6 & six $b$ elements & $E$   & ${8}\choose{6}$ =28 classes \\
7 & seven $b$ elements & $E$   & ${8}\choose{7} $=8  classes\\
8 & eight $b$ elements & $E$   & ${8}\choose{8}$ =1  classes \\
9 & $O, \overline{M}_1$ & $E, \overline{M}_5$   & 7 classes \\
10 & $O$ & $E$   &  ${7}\choose{1}$ =1  classes \\
11 & $M_i$ & $E$   &  ${6}\choose{1}$ =6  classes \\
12 & $\overline{O}, \overline{M}_2,$ & $E, M_6,$ &   \\
 & $M_3, M_5$ & $\overline{M}_4, \overline{M}_1$   &  105  classes \\
13 & $M_1, M_2,$ & $\overline{M}_5, \overline{M}_6,$ &  \\
 & $M_5, M_4,$ & $\overline{M}_1, \overline{M}_3,$ & \\
 & $\overline{M}_1, \overline{M}_2,$ & $M_5, M_6,$ &  \\
 & $\overline{M}_5, \overline{M}_4$ & $M_1, M_3$    & 15 classes \\ 
 14 & $\overline{O}, \overline{E},  M_3,  M_4$ 
 & $O, E, \overline{M_3}, \overline{M_4}$ & 3 classes \\
  15 & None & $O, \overline{O}$, $\overline{E} $ & 1 class \\
   16 & $O, \overline{O}$ & $\overline{E} $ & 1 class \\
 17 & None & $M_1, M_2, M_3$ &  \\ 
 & & $\overline{M_1}, \overline{M_2}, \overline{M_3}$ & 1 class \\
....& & \\  \hline
\end{tabular}
\end{center}
\end{table}%

When we expand the velocity and magnetic fields using the Fourier basis function $\exp(i{\bf k} \cdot {\bf r})$, 
the nonlinear terms of Eqs.~(\ref{eq:u_k},\ref{eq:b_k})  are sums of quadratic products of the modes. For the time being, we ignore the forcing and focus on the symmetry properties of the product terms.  When we focus on  a single triad with wavenumber $({\bf k,p,q})$ satisfying a constraint ${\bf k= p+q} $, the structures of the nonlinear terms of the MHD equations are   
\begin{eqnarray} 
\partial_t \hat{u}(\mathbf k)  & \sim & \hat{u}(\mathbf p)  \hat{u}(\mathbf q) + \hat{b}(\mathbf p)  \hat{b}(\mathbf q) ,\label{eq:unit_triad_u}  \\
\partial_t \hat{b}(\mathbf k) & \sim & \hat{u}(\mathbf p)  \hat{b}(\mathbf q) + \hat{b}(\mathbf p)  \hat{u}(\mathbf q) .
\label{eq:unit_triad_b}
\end{eqnarray}
If we denote $k_i = n_i$, $p_i =  l_i$, and $q_i = m_i$ with $i=(x,y,z)$ and $l_i, m_i, n_i$ as integers, then the  condition ${\bf k= p+q} $ yields $(n_x,n_y,n_z) =  (l_x +m_x, l_y +m_y, l_z+m_z)$.  Here we assume the box size to be $(2\pi)^3$.  For the free-slip and insulating boundary conditions, we employ basis functions involving $\sin({\bf k \cdot r})$ and $\cos({\bf k \cdot r})$.  Since $2 i \sin({\bf k \cdot r}) = \exp({\bf k \cdot r}) -  \exp(-{\bf k \cdot r})$ and $2\cos({\bf k \cdot r} =\exp({\bf k \cdot r}) -  \exp(-{\bf k \cdot r})$, the product rule indicates that $(n_x,n_y,n_z)$ could be one of the  following: $(\pm l_x \pm m_x, \pm l_y  \pm m_y, \pm l_z \pm m_z)$, depending on their forms ($\sin$ or $\cos$). 

To make a connection with the Klein group, it is convenient to represent the velocity Fourier mode $u_i(p_x, p_y, p_z)$  using $(e~P(p_x)~P(p_x)~P(p_x))$, and $b_i(q_x, q_y, q_z)$ modes using $(o~P(q_x)~P(q_x)~P(q_x))$, where $P$ is the parity operator.  It is important to note that the forcing function ${\bf f}$ of Eq.~(\ref{eq:force}) is at ${\bf k} = (1,1,1)$ and it forces the velocity field,  hence it   belongs to the $(eooo)$ category.

Using the rules of addition, even+even = even, even+odd = odd, and odd+odd = even, we obtain the product rules described in Table~\ref{tab:3d_product_rule}. Here we list $A \times B =C$, where the 16 elements of $A$ are listed in the first column, while the 16 elements of $B$ are  listed in the first row.  The first 8 entries correspond to the velocity field, while the latter 8 to the magnetic field. In terms of parity, the 16 elements of rows or columns in increasing order are $(eeee)$, $(eooo)$, $(eeoo)$, $(eoeo)$, $(eooe)$, $(eeeo)$, $(eoee)$, $(eeoe)$, $(oeee)$, $(oooo)$, $(oeoo)$, $(ooeo)$, $(oooe)$, $(oeeo)$, $(ooee)$, $(oeoe)$.  The result $C$ makes the table. To illustrate, according to Eq.~(\ref{eq:unit_triad_b}), $\hat{\bf u}(1,1,1) \times  \hat{\bf b}(3,1,3)$ contributes to  $\partial \hat{\bf b}(4,2,4)/\partial t$; this product is listed as $O \times \bar{O} = \bar{E}$ (see multiplication of row 3 with column 11).  Similarly, $\hat{\bf u}(2,1,1) \times  \hat{\bf b}(3,1,3)$ contributes to $\partial \hat{\bf b}(5,2,4)/\partial t$ is captured by $M_1 \times \bar{O} = \bar{M}_5$ (see multiplication of row 4 with column 11).  Note that  $\hat{\bf u}(1,1,1)   \in O$, $\hat{\bf u}(2,1,1)   \in M_1$, $\hat{\bf b}(5,2,4) \in M_5$, etc.  Some of the other examples in the table are  $O\times O = E$, $O\times M_1 = M_5$,   etc.  

The multiplication table Table~\ref{tab:3d_product_rule} is divided into four subparts.  The first quadrant involves multiplication of the velocity modes only, and it yields a velocity mode due to ${\bf u} {\bf u} \rightarrow {\bf u}$ [see Eq.~(\ref{eq:unit_triad_u})].  The last quadrant involves products of two magnetic modes, and it yields a velocity mode  since ${\bf b} {\bf b} \rightarrow {\bf u}$.  The second and third quadrants deal with products of a velocity mode and a magnetic mode, and the resultant field is a magnetic mode following the multiplication rules $ {\bf u} {\bf b} \rightarrow {\bf b}; {\bf b} {\bf u} \rightarrow {\bf b}$.  Interestingly it is captured quite nicely by the 4-index representation of ${\bf u,b}$ as  $(e~p_x~p_y~p_z)$  and $(o~q_x~q_y ~q_z)$ respectively.  Hence, the 16 elements of Table~\ref{tab:3d_product_rule} form a Klein 16-group $Z_2 \times Z_2 \times Z_2 \times Z_2$.   The multiplication table shows that the group is abelian.  Also, the elements of the first quadrant (the velocity modes) form a subgroup $Z_2 \times Z_2 \times Z_2$, which is a Klein eight-group.  In addition, it is easy to show that the usual symmetry of MHD, ${\bf u} \rightarrow {\bf u}$ and ${\bf b} \rightarrow -{\bf b}$,  is trivially satisfied by the multiplication table; here, all the $\hat{b}$ modes change sign, but  all the $\hat{u}$ modes retain their sign.

An interesting feature  of the above multiplication Table is $O \times M_1 = M_5$, $O \times M_5 = M_1$, and $M_5 \times M_1 = O$; due to the above, $M_1$ and $M_5$ are termed as complement of each other, i.e., $M_1' = M_5$ and $M_5' = M_1$.  Similarly $M_2$ and $M_6$ are complements of each other, so are $M_3$ and $M_4$.  Similar rules apply for magnetic modes as well. Another feature to note is that $E$ is the identity of the group because $E\times X = X$, where $X$ stands for any element of the group.  From the product rule, it is evident that $E$ cannot change sign under  reversal. Also note that $\overline{E}$ could reverse under reversal (to be described below).   

In a dynamo, the Fourier modes usually fluctuate around their average values, which could be finite or zero.  Some of these modes switch sign after a reversal, while  some do not; these modes are the reversing (R) and non-reversing modes (NR) modes respectively.  The R and  NR modes are determined using the rules of the multiplication table.  In Table~\ref{tab:MHD_rules}, we list only some of these classes of such modes since there are just too many entries to be comprehensively listed.  For illustration, we assume that the nonzero modes of our system are $E$ and  $\overline{E}$ only, and the rest of the modes are negligible, then we have two possibilities---(a) the  modes of $E$ and  $\overline{E}$ classes (with nonzero amplitudes) do not reverse (NR), or (b) the modes of $\overline{E}$ class reverse, but those of $E$ class do not reverse. In general, suppose $X$ modes do not switch sign (column 3 of Table 2), $Y$ switch sign (column 2 of Table 2), and $Z$ take small values denoted by $\epsilon$ (modes not covered in columns 2 and 3 of Table 2).  Then, from the dynamical equations and the product rules of Table I, we can deduce that if $\{X, Y, Z\}$ is a solution of the equations, then  $\{X, -Y, Z\}$ is also a solution of the equations.

The entries of Table II are constructed using the multiplication rules of Table I.  As described earlier, $E$ is the identity element of the group, and it does not change sign.  Since $O\times E =O$ and $O \times O = E$, we deduce that  $O$ can change sign, which is the entry of item 10 of Table II.  Similar arguments work if we replace $O$ by any of  $M_i$'s, which yields the item 11 of Table II; note that there are 6 possibilities of choosing an $M_i$.  The products rules $\bar{O} \times \bar{O} =E$, $\bar{O} \times \bar{M}_2 = M_6$, $\bar{M}_2 \times \bar{M}_2 =E$ yields the item 12 of Table II. Similar analysis yields other entries of  Table II. 

Now let us bring in the effects of external force.  We rewrite Eqs.~(\ref{eq:unit_triad_u}, \ref{eq:unit_triad_b}) with the external force ${\bf f}$ on the velocity field:
\begin{eqnarray} 
\partial_t \hat{u}(\mathbf k)  & \sim & \hat{u}(\mathbf p)  \hat{u}(\mathbf q) + \hat{b}(\mathbf p)  \hat{b}(\mathbf q) + \hat{f}(\mathbf k) ,\label{eq:unit_triad_uf}  \\
\partial_t \hat{b}(\mathbf k) & \sim & \hat{u}(\mathbf p)  \hat{b}(\mathbf q) + \hat{b}(\mathbf p)  \hat{u}(\mathbf q) .
\label{eq:unit_triad_bf}
\end{eqnarray}
If the Fourier mode $\hat{f}(\mathbf k) $ does not change sign, then $\hat{u}(\mathbf k)$ cannot change sign to preserve the parity of Eq.~(\ref{eq:unit_triad_uf}).  Alternatively, if $\hat{f}(\mathbf k) $ changes sign during an event, then $\hat{u}(\mathbf k)$  would change sign too.  Thus, external force too plays an important role in determining reversing and non-reversing modes.

The aforementioned discrete symmetry is useful for understanding dynamo reversals.  For example, researchers could test whether the quadrupolar mode of geodynamo reverses or not.  Also, the real-space signature of the magnetic field before and after a reversal could reveal important and interesting clues about the system dynamics.  

In the next section we generalise the above arguments to magneto-convection.

\section{Extension of symmetry arguments to magneto-convection}
\label{sec:magneto_convection}
Geodynamo and solar dynamo are driven by convection.  Hence, researchers model such equations using magneto-convection whose equations are~\cite{Roberts:RMP2000,Morin:PF2004}
\begin{eqnarray}
\frac{\partial {\bf u}}{\partial t} + ({\bf u} \cdot \nabla){\bf u} & = & - \nabla P + ({\bf j \times b}) +  \alpha g T \hat{z} + \nu \nabla^2 {\bf u}, \label{eq:u_MHDC} \\
\frac{\partial \mathbf{b}}{\partial t} + (\mathbf{u} \cdot \nabla)\mathbf{b} & = & (\mathbf{b} \cdot \nabla)\mathbf{u} + \eta \nabla^2 \mathbf{b}, \label{eq:b_MHDC}\\
\frac{\partial T}{\partial t} + ({\bf u} \cdot \nabla) T & = &   \kappa \nabla^2 T, \label{eq:thB} \\
\nabla \cdot {\bf u} & = & \nabla \cdot {\bf b} = 0, \label{eq:delUB_zero}
\end{eqnarray}
where $\mathbf{u,b}$ are the velocity and magnetic fields, $T$ is the temperature field,  ${\bf j}$ is the current density, $P$ is the pressure field, $\nu,\eta,\kappa,\alpha$ are respectively the kinematic viscosity, magnetic diffusivity, thermal diffusivity, and thermal expansion coefficient of the fluid, and $-g \hat{z}$ is the acceleration due to gravity.  

When we compare the aforementioned equations with those of MHD (\ref{eq:u_non}-\ref{eq:inc}), we observe that the magneto-convective systems are forced externally using buoyancy $\alpha g T \hat{z}$.  Also, magneto-convection has an additional nonlinear term, $ ({\bf u} \cdot \nabla) T$.  In terms of interactions among the Fourier modes, the terms  $ ({\bf u} \cdot \nabla) T$ and  $ ({\bf u} \cdot \nabla) {\bf u}$ have some structure, hence the Fourier modes of the  ${\bf u}$ and $T$ fields belong to the same class, i.e. $E, O, M_1, M_2, M_3, M_4, M_5, M_6$.  These arguments are  in the same lines as those of Chandra and Verma~\cite{Chandra:PRL2013} and Verma {\em et al.}~\cite{Verma:POF2015}, which are described in Appendix A.  Thus, the group-theoretic structure of magneto-convection is same as that of MHD, i.e. Table~\ref{tab:3d_product_rule}.  Also, the classes of reversing and non-reversing modes modes for MHD and magneto-convection should be the same.  However, an important point to remember is that the temperature field $T$ drives the flow, hence the symmetries of $\hat{T}$ will dictate the symmetry classes of ${\bf u}$. 

 In the next section, we demonstrate one of the classes of Table~\ref{tab:MHD_rules} using a DNS of a reversal in a Taylor-Green dynamo. 

\section{A demonstration of the group-theoretic model using a dynamo DNS}\label{sec:DNS}
We perform numerical simulations of a dynamo reversal using a pseudo-spectral solver TARANG.~\cite{Verma:Pramana2013}    We use the fourth-order Runge-Kutta (RK4) scheme for time advancement, Courant-Friedrichs-Lewy (CFL) condition for choosing the variable time step, and 2/3 rule for dealiasing. For our simulations, we choose a box of dimesion $\pi^3$ with resolution of $128^3$ grid points. We employ a free-slip boundary condition for the velocity field and insulating boundary condition for the magnetic field at all the walls.   For the same, we use the basis functions of Eqs.~(\ref{eq:ux_3d}-\ref{eq:bz_3d}).  We implement these basis functions by applying appropriate symmetries  on a periodic box of $(2\pi)^3$ size (see Sec.~II).  The box satisfying free-slip and insulating boundary conditions is in the first quadrant of the $(2\pi)^3$ cube.  In this paper we use nondimensionalised time $L_0/U_0$, where $L_0,U_0$ are the length and velocity scale of the system. Note that our scheme is same as that of  Krstulovic {\em et al.}~\cite{Krstulovic:PRE2011}.  Also see \cite{Chen:JFM2015}.

\begin{figure}
\begin{center}
\includegraphics[scale=1.0]{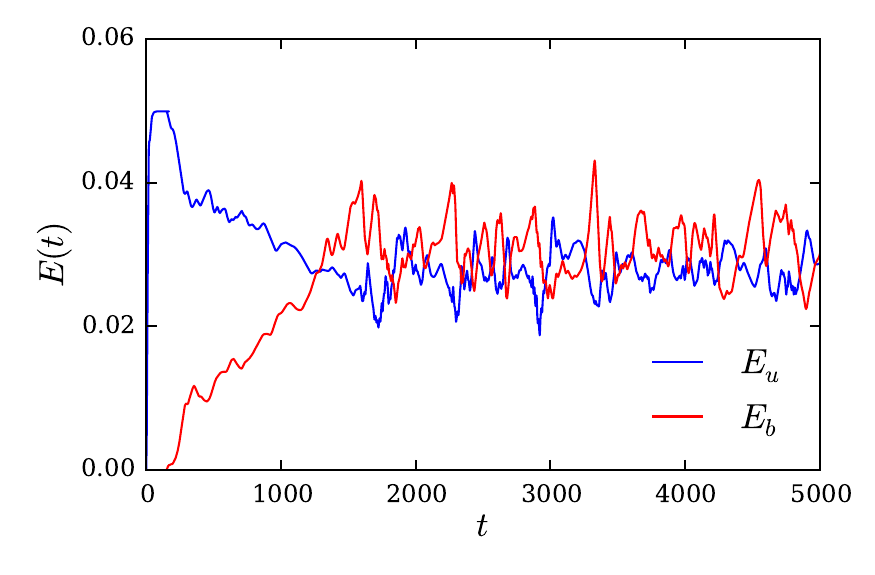}
\caption{Evolution of the total kinetic energy $(E_u)$ and total magnetic energy ($E_b$) in the dynamo simulation. As initial condition, a seed magnetic field is injected into the steady state of the fluid flow.  The magnetic energy grows exponentially and  reaches a steady state. Time in this plot and subsequent plots are in nondimensionalised units.}
\label{fig:energy}
\end{center}
\end{figure}

The Taylor-Green vortex (Eq.~(\ref{eq:force}) with ${\bf v}$ as the field) was used as the initial condition for the velocity field. A small spectrally band-limited $(k = 2, 4)$ field was chosen as the initial condition for the magnetic field.  We ran several sets of simulations for various forcing amplitudes $F_0$, viscosity $\nu$, magnetic diffusivity $\eta$. We achieved a sustained dynamo for $\nu = 0.01$, $\eta = 0.001$, $F_0 = 0.05$. The Prandtl number $\mathrm{Pm} = \nu/\eta = 10$, while the Reynolds number $\mathrm{Re} = UL/\nu = 70$, and magnetic Reynolds number $\mathrm{Rm}=  UL/\eta =700$, where $U,L$ are the large-scale velocity and length of the flow.  Thus our flow is not turbulent, but $\mathrm{Rm}$ is large enough to sustain a dynamo.  In all our simulations, $k_{max}{\color{blue}{\zeta}}$ (where $k_{max} = 64$ is the maximum wavenumber, and ${\color{blue}{\zeta}}$ is the Kolmogorov's length scale) is always greater than 2, thus our simulation is well resolved.  We observe that during the steady-sate, the kinetic and magnetic energies are equipartitioned and they fluctuate around their mean  (see Fig.~\ref{fig:energy}). Although Prandtl number $\mathrm{Pm} = 10$ for our simulation is not very large, some of the conclusions drawn here may hold for large $\mathrm{Pm}$ or for $\mathrm{Pm} \rightarrow \infty$.   Schober {\em et al.}~\cite{Schober:PRE2015} gave analytical calculations to predict the ratio of initial kinetic energy to the magnetic energy at saturation in the $\mathrm{Pm} \gg 1$ limit. The fraction of initial turbulent kinetic energy that is converted into magnetic energy at saturation is around $40\%$ for incompressible flows, as reported in previous investigations ~\cite{Haugen:PRE2004, Federrath:PRL2011, Schober:PRE2015}. For our simulation, the ratio is $0.03/0.05 \approx 60\%$, which is not very far from the results in $\mathrm{Pm} \gg 1$ case given that the limit is not strictly applicable here.  Alexakis~\cite{Alexakis:PRE2011} investigate nonlinear dynamos in the $\mathrm{Pm}=\infty$ limit.  The flow exhibits a weak increase in the magnetic energy as $\mathrm{Rm}$ is increased.

\begin{figure}
\begin{center}
\includegraphics[scale=0.7]{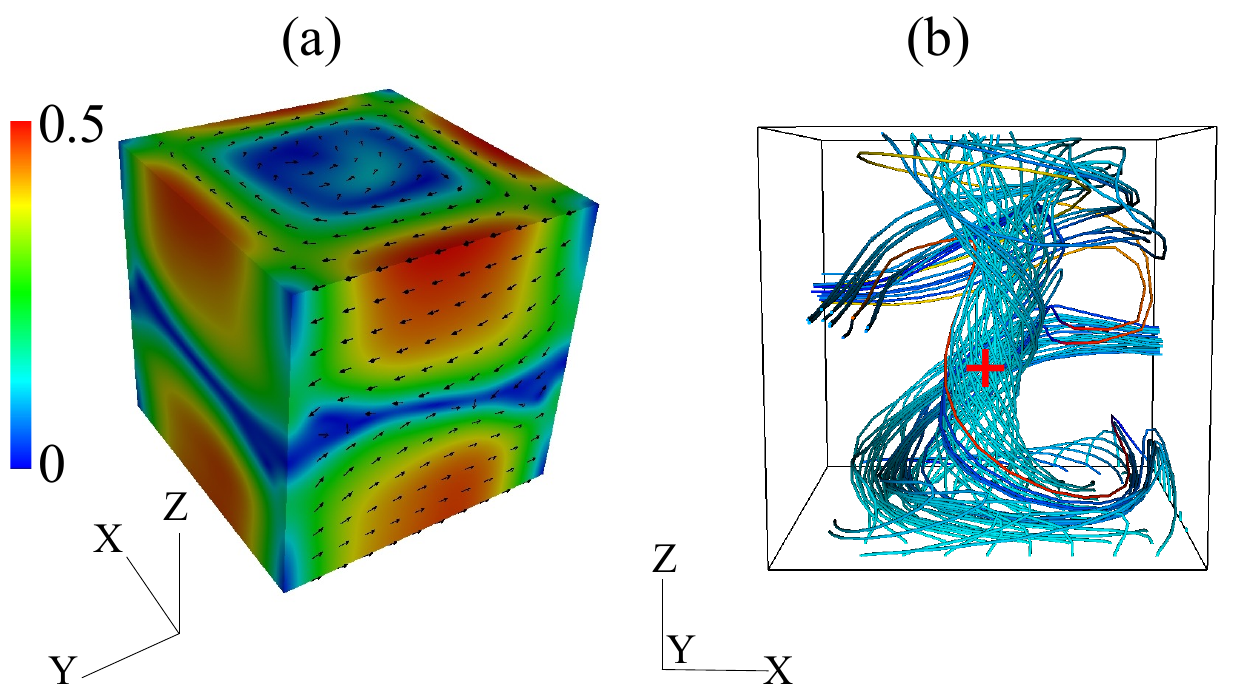}
\caption{Snapshots of the velocity and magnetic fields in the steady state of our dynamo simulation. (a) The vector plot and the density plot of the speed of the flow with blue color depicting slow flows, and red color the fast flows.   (b) Plot of the magnetic field lines in which the axial dipolar structure is clearly visible.  A real space probe was placed at the grid point  $(64,90,64)$ (shown as red + sign in (b)) exhibits reversals of $u_x$.}
\label{fig:real_space}
\end{center}
\end{figure}

Figure~\ref{fig:real_space}(a) exhibits a snapshot of the velocity field in the steady state; here the blue and red colors represent the regions with small and large speeds $|{\bf u}|$ respectively.   The figure shows a shear layer in the middle of the box that separates the two counter-rotating eddies (at the bottom and top of the box); this flow structure closely resembles that in the VKS experiment.  Figure~\ref{fig:real_space}(b), illustrating the magnetic field lines, indicates  an axial dipole oriented along the z axis.  

\begin{figure}[htp]
\begin{center}
\includegraphics[scale=0.32]{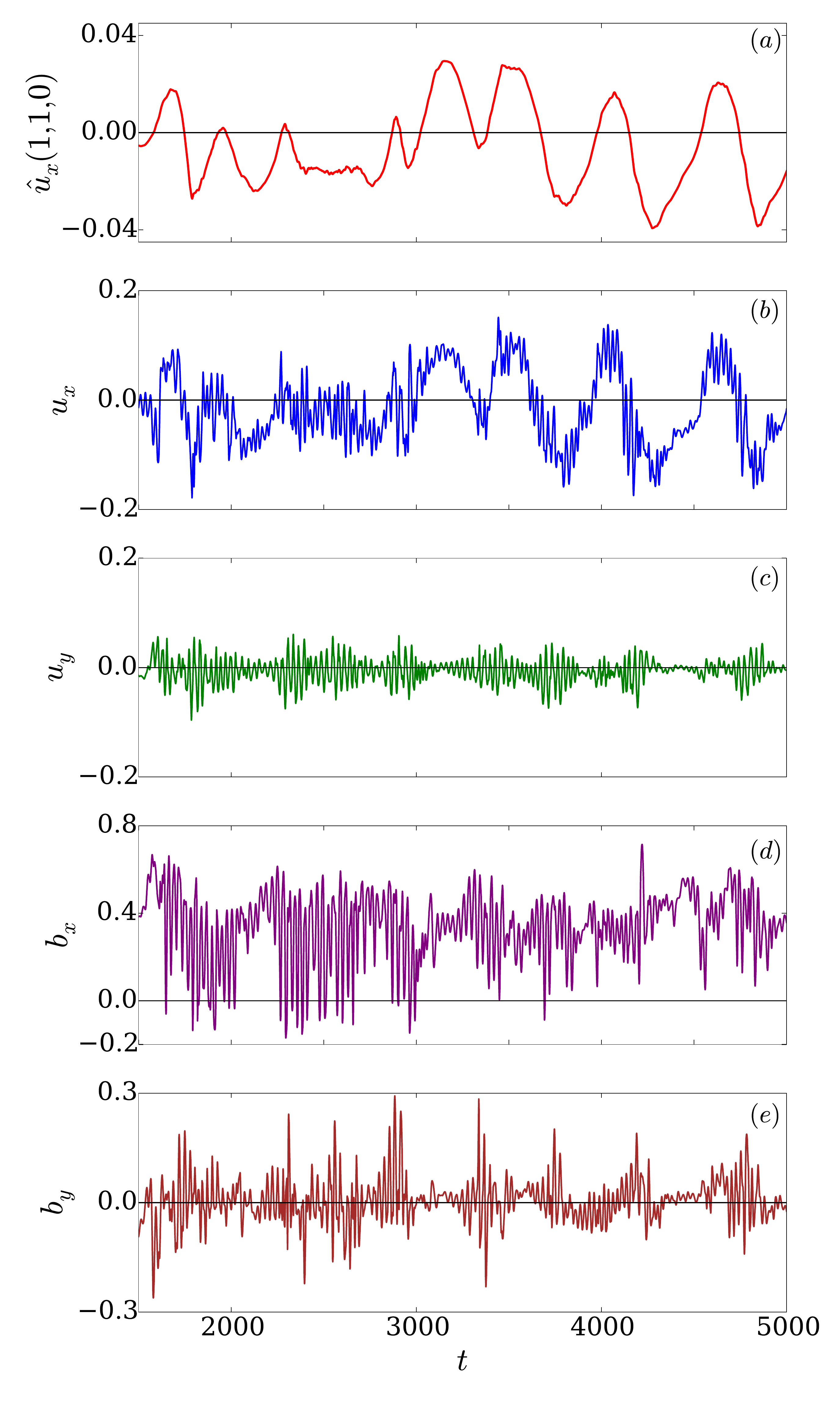}
\caption{Plot of the time series of (a) the Fourier mode $\hat{u}_x(1,1,0)$. Plots of the time series of real space (b) $u_x$, (c) $u_y$, (d) $b_x$, (e) $b_y$ probed at the grid point $(64,90,64)$ (see Fig.~\ref{fig:real_space}(b)).}
\label{fig:64_90_64}
\end{center}
\end{figure}

We ran our simulation for 5000 eddy turnover time during which the axial magnetic field does not exhibit any field reversal.  However, the velocity field component $u_x$ exhibits reversals near the grid point $(64,90,64)$, as shown in the time series plot of Fig.~\ref{fig:64_90_64}(b).  In  Fig.~\ref{fig:64_90_64}(a) we exhibit the time series of the Fourier mode $\hat{u}_x(1,1,1)$ that shows reversal in sync with those of Fig.~\ref{fig:64_90_64}(b).  Interestingly,  $b_x$ retains its sign, while $u_y$ and $b_y$ fluctuate around zero, as shown in the time series plots of Fig.~\ref{fig:64_90_64}(c,d,e).   The aforementioned phenomena indicates an interesting dynamics that may become  apparent using the properties of the reversing and non-reversing Fourier modes.

\begin{figure}
\begin{center}
\includegraphics[scale=0.32]{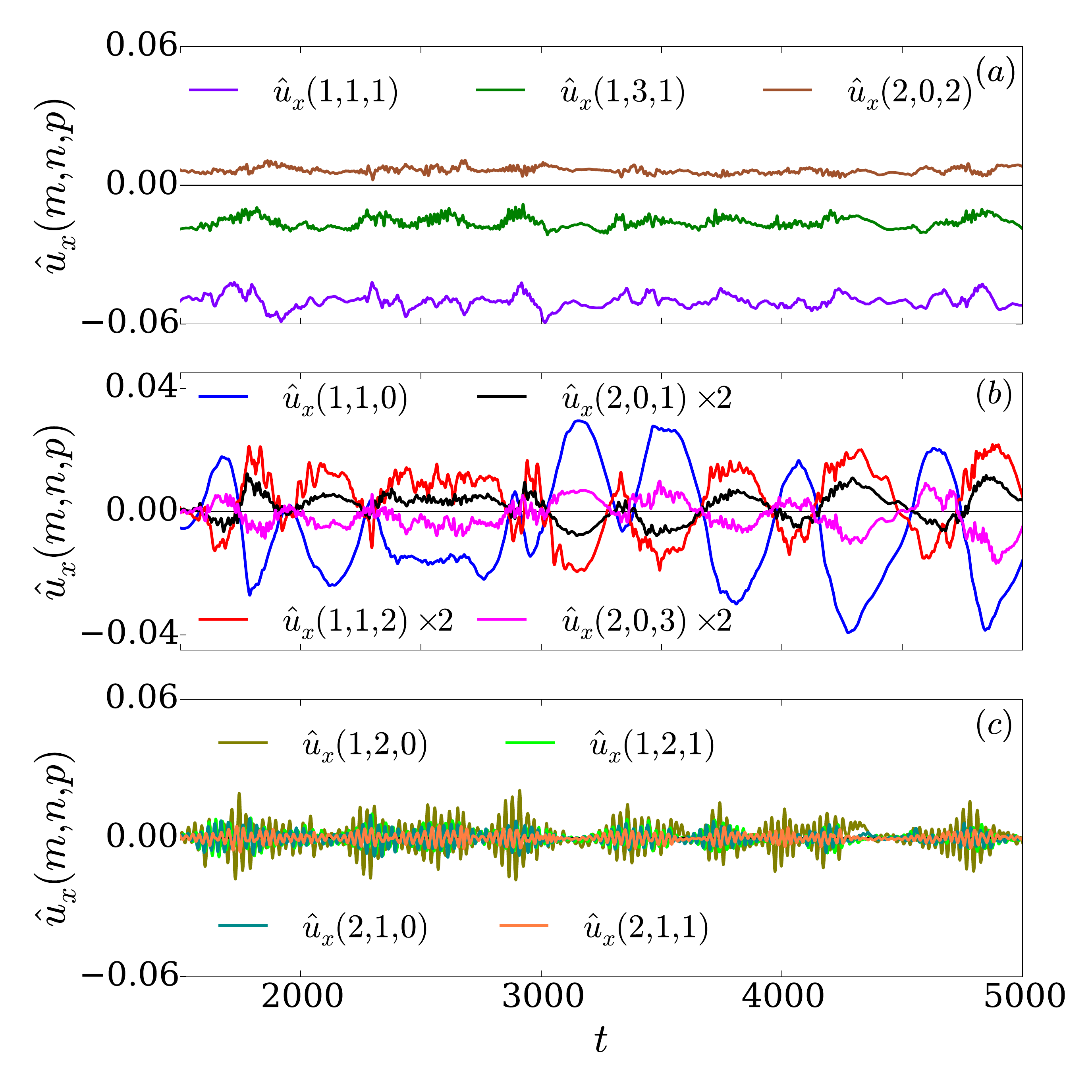}
\caption{Plots of the time series of the amplitudes of some of the dominant velocity Fourier modes  that are (a) non-reversing, (b) reversing, and (c) fluctuating modes.}
\label{fig:u_x}
\end{center}
\end{figure}

In Fig.~\ref{fig:u_x}, we plot the time series of the amplitudes of the dominant velocity modes.  Figure~\ref{fig:u_x}(a,b,c) exhibit the non-reversing, reversing, and vanishing modes respectively.  We observe that the modes $\{ (1,1,1), (1, 3,1) \} \in O$ and $(2,0,2) \in E$ are non-reversing.  The modes $\{  (1,1,0), (1,1,2) \}  \in M_3$ and $\{ (2,0,1), (2,0,3)\}   \in M_4 $ reverse.  All other modes fluctuate around zero.  From these observations, we conclude that the modes in $O, E$ classes are non-reversing, and the modes in $M_3, M_4$ are reversing, and rest all are vanishing.  

We can explain the above features using the symmetry classes of Sec.~\ref{sec:Sym} as follows: From the rules of Table~\ref{tab:3d_product_rule}, the $E$ modes cannot flip. A {\em constant-amplitude} forcing that belongs to $(ooo)$ class is employed to the momentum equation. As a result, the velocity modes of class $O$ do not switch sign, as argued in Sec.~\ref{sec:Sym} while discussing Eq.~(\ref{eq:unit_triad_uf}). Mishra {\em et al.}~\cite{Mishra:PRE2015} observed similar behaviour in the reversals of  Kolmogorov flow; they observed that the constant forcing at (6,6) makes the velocity Fourier mode $\hat{\bf u}(6,6)$ non-reversing. Note that modes $\{{\bf u}(2,0,2), {\bf u}(1,1,0), {\bf u}(1,1,2)\}$ form a triad, and they respect the rules of Table~\ref{tab:3d_product_rule}.

\begin{figure}[htp]
\begin{center}
\includegraphics[scale=0.32]{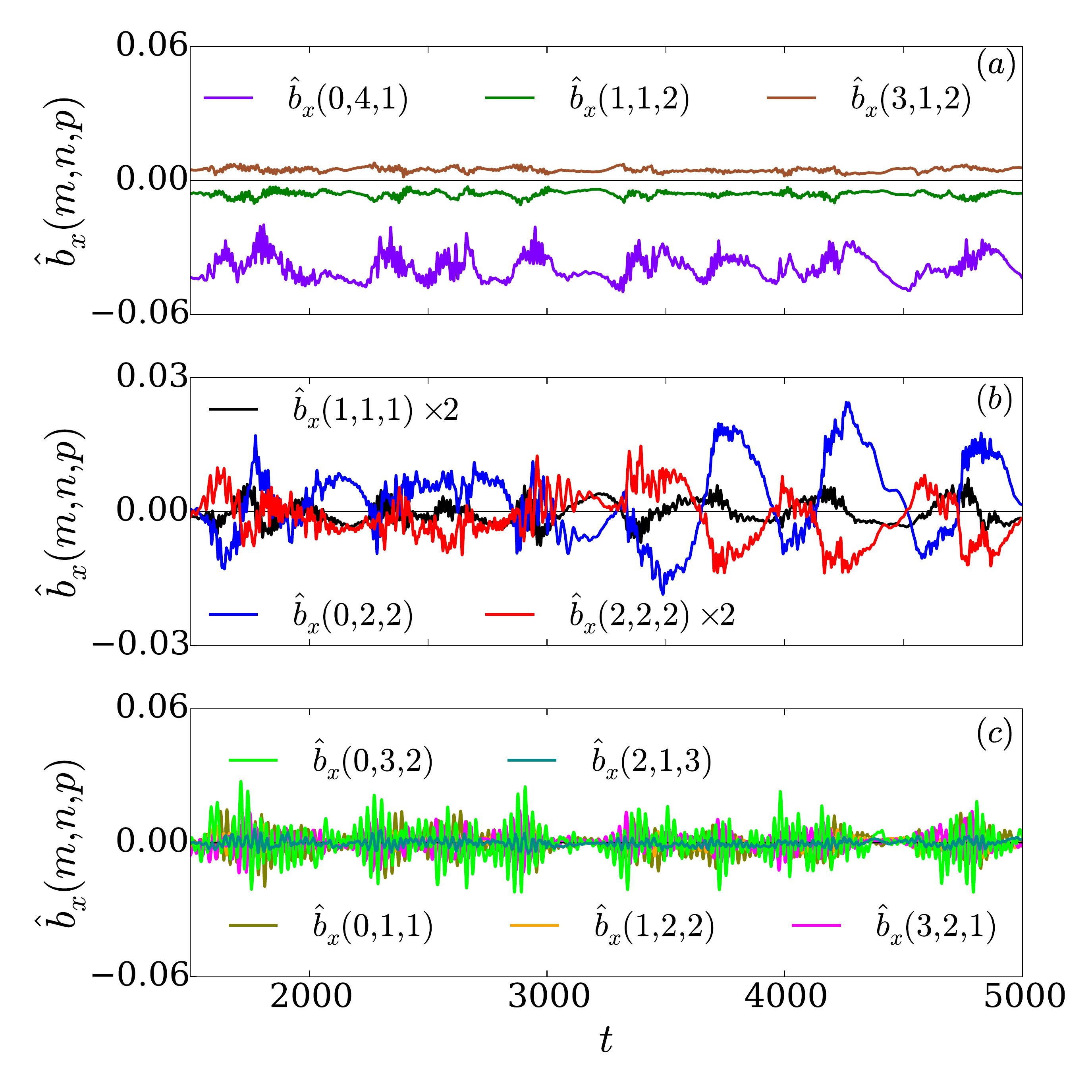}
\caption{Plots of the time series of the amplitudes of some of the dominant magnetic Fourier modes  that are (a) non-reversing, (b) reversing, and (c) fluctuating modes. }
\label{fig:b_x}
\end{center}
\end{figure}

Now let us study the   magnetic modes.  Figure~\ref{fig:b_x} exhibits the times series of the dominant magnetic modes, and it shows that the modes in $\bar{O}, \bar{E}$ classes are reversing, and the modes in $\bar{M}_3, \bar{M}_4$ are non-reversing, and rest all are vanishing.  Some of the important magnetic modes are---$\{ (1,1,1) \} \in O$ (reversing), $\{ (0,2,2),(2,2,2) \} \in E$ (reversing), $\{  (1,1,2), (3,1,2) \}  \in M_3$ (non-reversing), and $\{ (0,4,1)\}   \in M_4 $ (non-reversing).  Two {\bf b} modes interact nonlinearly with one of the {\bf u} modes.  Some of the interacting triads are $\{{\bf b}(0,2,2), {\bf b}(1,1,1), {\bf u}(1,1,1)\}$, $\{{\bf b}(0,2,2), {\bf b}(1,1,2), {\bf u}(1,1,0)\}$, and  $\{{\bf b}(1,1,2), {\bf b}(1,1,1), {\bf u}(2,0,1)\}$. It is easy to verify that all these triads satisfy the multiplication rules of Table~\ref{tab:3d_product_rule}.   In summary, in the dynamo reversals of our DNS, the modes belonging to the class $\{ O, E, \overline{M}_3, \overline{M}_4 \}$ do not reverse, those in the class $\{ \overline{O}, \overline{E}, {M}_3, {M}_4 \}$ reverse, and the remaining ones are vanishing ($\epsilon$) and they fluctuate around zero.  This is item (14) in Table~\ref{tab:MHD_rules}.

 In our dynamo simulation, the dipolar component of the magnetic field does not reverse. It is possibly due to the constant forcing term ${\bf f}(1,1,1)$; it is connected to the non-reversal regime of the VKS experiment when the propellers are rotated with equal and opposite frequencies~\cite{Gissinger:EPL2010}.   Gissinger {\em et al.}~\cite{Gissinger:EPL2010} obtained reversals when the symmetry of the  ${\bf f}(1,1,1)$ mode was broken.  We believe similar scheme, for example, randomly-varying ${\bf f}(1,1,1)$ could induce reversals in the dipolar magnetic component.
 
We revisit the reversals of Fig.~\ref{fig:64_90_64}.  The reversal of $u_x$ at the point $(64,90,64)$ is a combined effect of all the velocity Fourier modes.  At this point, the effects of the non-reversing modes appear to cancel each other, while those of the reversing modes add up.  We are studying the reconstruction of the real-space time series using the dominant modes, and it will be reported in a future communication.  These observations indicate that the time series of the signals in real space and Fourier space complement each other, and they need to be studied carefully.  Such analysis may yield interesting insights into the velocity and magnetic field reversals in solar and geo dynamo. 

In the next section we will describe some other dynamo reversals and identify their category in Table~\ref{tab:MHD_rules}.

\section{Connections with earlier dynamos}
\label{sec:earlier_dynamos}
Regarding Gissinger {\em et al.}'s dynamo model~\cite{Gissinger:EPL2010},  it is difficult to relate the $D, Q$, and $V$ variables of Eqs.~(\ref{eq:D_eqn}-\ref{eq:qvd_eqn}) to the Fourier modes of a Cartesian box.   Qualitatively we could argue that $D \rightarrow \overline{O}$,  $Q \rightarrow \overline{E}$, and $V \rightarrow {O}$, for which $O$ and $ \overline{O}$ would reverse, but $ \overline{E}$ will not reverse, as indicated by item (16) of Table~\ref{tab:MHD_rules}.  

In VKS experiment, the magnetic field does not reverse when the propellers rotate with equal and opposite frequencies~\cite{Berhanu:EPL2007}. This feature corresponds to  a constant and  dipolar $V$.  This property follows from the symmetry property item (15) of Table~\ref{tab:MHD_rules}.  The magnetic field reversals occur when the two propellers rotate with unequal frequencies that may correspond to the excitation of mixed modes $M_i$.  Such configuration may correspond to item (12) or its variations in Table~\ref{tab:MHD_rules}.

Verma {\em et al.}~\cite{Verma:PRE2008} constructed a six-mode dynamo model that does not exhibit reversal.  In fact, it does not exhibit any time-dependent behaviour, which has been a puzzle.  However, we can understand this phenomenon using the symmetries.  In Verma {\em et al.}'s model~\cite{Verma:PRE2008}, the participating modes are ${\bf u}(1,0,1), {\bf u}(0,1,1), {\bf u}(1,1,2)$, and ${\bf b}(1,0,1), {\bf b}(0,1,1), {\bf b}(1,1,2)$ of which ${\bf b}(1,1,2) = 0$. The velocity fields of the model is forced by constant ${\bf f}_1(1,0,1)$ and $ {\bf f}_2(0,1,1)$. Clearly these modes belong to the classes $M_1, M_2, M_3, \overline{M}_1, \overline{M}_2, \overline{M}_3$.  Due to the constant forcing and symmetry properties, ${\bf u}(1,0,1), {\bf u}(0,1,1)$ modes that belong to $M_1, M_2$ classes respectively cannot flip at all.  The mode ${\bf u}(1,1,2) \in M_3$ does not flip due to symmetry properties.  It is possible that both ${\bf b}(1,0,1) \in \overline{M}_1, {\bf b}(0,1,1) \in \overline{M}_2$ flip, but it appears that non-reversing velocity modes inhibits reversals of the magnetic modes (see item (17) of Table~\ref{tab:MHD_rules}).  This is how we can provide a qualitative explanation for the  the non-reversing behaviour of the model proposed by Verma {\em et al.}~\cite{Verma:PRE2008}.

Yanagisawa {\em et al.}~\cite{Yanagisawa:PRE2010} study flow reversals in magneto-convection in Cartesian geometry.  It will be interesting to study the symmetry classes of the reversals in such systems.  In addition, many dynamo models involve magneto-convection in spherical geometry. The symmetry class of such systems are more general than those presented in the paper.  We need to generalize the analysis of Sec.~\ref{sec:Sym} to spherical geometry that involves continuous symmetry.

\section{Conclusions and Discussions}\label{sec:Con}
In this paper, we have investigated the properties of dynamo reversals in box geometry using the properties of nonlinear interactions among the Fourier modes. The Fourier basis are convenient for studying the triadic interaction ${\bf k} = {\bf p} + {\bf q}$.  As $\{ \mathbf{u} \rightarrow \mathbf{u}, \mathbf{b} \rightarrow -\mathbf{b} \}$ is a symmetry of the MHD equations (Eq.~(\ref{eq:u_non})$-$(\ref{eq:b_gauss})), so it is generally argued that the signs of all the magnetic Fourier modes change simultaneously.  As argued in this paper, this is not the case.   We present symmetry arguments to derive the reversing and non-reversing Fourier modes.  We show that the modes $\{ E \}, \{ O \}, \{ M_1 \}, \{ M_2 \}, \{ M_3 \}, \{ M_4 \}, \{ M_5 \}, \{ M_6 \},\\ \{ \overline{E} \}, \{ \overline{O} \}, \{ \overline{M_1} \}, \{ \overline{M_2} \}, \{ \overline{M_3} \}, \{ \overline{M_4} \}, \{ \overline{M_5} \}, \{ \overline{M_6} \}$ form an abelian Klein-16 group $Z_2 \times Z_2 \times Z_2 \times Z_2$.  We show that the even Fourier modes of velocity field, belonging to the class $\{ E \}$, do not switch sign because it is the identity element of the group. On the other hand, the even Fourier mode of magnetic field belonging to class $\{ \overline{E} \}$ does not have any such constraint, and they can switch sign.  Our arguments show that the reversing and non-reversing modes can come in various combinations, some of which are listed in Table~\ref{tab:MHD_rules}.  We also generalise the symmetry arguments to magneto-convection. 

We perform a DNS that exhibits reversals in a Taylor-Green flow, and study the reversing and non-reversing modes during a reversal.   We observe that the modes belong to the classes $\{ O, E, \overline{M}_3, \overline{M}_4 \}$ do not reverse, those in classes $\{ \overline{O}, \overline{E}, {M}_3, {M}_4 \}$ reverse, and the remaining ones are vanishing ($\epsilon$).  Interestingly, in the real space, the dipolar magnetic field as well as the large-scale velocity field do not flip, but we observe reversals of the velocity field at one of the points in the real space.  Thus, probing reversing and non-reversing Fourier modes provide a very useful diagnostics for dynamo reversals.  These kind of studies may prove particularly useful for the solar and geo-dynamo.
 
We remark that the symmetry arguments presented in the paper are similar to those of ~\cite{Weiss:JStatPhys1985, Jennings:MNRAS1991, Knobloch:MNRAS1996, Knobloch:MNRAS1998}  and more recently, of P\'{e}trellis {\em et al.}~\cite{Petrelis:PRL2009} and Gallet {\em et al.}~\cite{Gallet:GAFD2012} who exploit the rotation and mirror symmetries of the equation.  Our arguments are group-theoretic and algebraic, hence the symmetry classes are easier to derive.   

The arguments presented in the paper are based on discrete symmetries.  Note that a cylinder has U(1) symmetry (rotation about the cylindrical axis), while a sphere has SO(3) rotation symmetry; these symmetries are continuous in nature.  Hence the arguments presented in the paper are not directly extensible to these systems; further work is required in this direction.  Note however that the spherical harmonics are used as the basis functions for studying dynamos in sphere, and some of the arguments presented here on discrete symmetries could be extended to the spherical harmonics.  These ideas may prove very useful for studying the reversal properties of multi-polar modes reported in spherical dynamos~\cite{Kutzner:PEP2002,Oruba:GRL2014}.  We hope such studies will be taken up in near future.

 The symmetry properties discussed in this paper are valid for MHD, and they are not necessary extendible to other systems that exhibit dynamo.  We need to extend our analysis for the same.

\section*{Acknowledgements}
We are thankful to Abhishek Kumar for extensive help in computation, and to Giorgio Krstulovic for fruitful discussions. We are grateful for to an anonymous referee for useful comments. The suggestions helped us improve the quality of the manuscript substantially. The computer simulations were performed on Shaheen II of the Supercomputing Laboratory at King Abdullah University of Science and Technology (KAUST) under the project K1052.  This work was supported by the Indo-French research project IFCPAR/CEFIPRA contract 4904-A, and by the Indo-Russian project (DST-RSF) project INT/RUS/RSF/P-03.

\section*{Appendix A: Symmetry classes  for the flow reversals in RBC}

The equations for Rayleigh-B\'{e}nard convection (RBC) are~\cite{Chandra:PRL2013}
\begin{eqnarray}
\frac{\partial {\bf u}}{\partial t} + ({\bf u} \cdot \nabla){\bf u} & = & - \nabla P + \alpha g T \hat{z} + \nu \nabla^2 {\bf u}, \label{eq:u} \\
\frac{\partial T}{\partial t} + ({\bf u} \cdot \nabla) T & = &   \kappa \nabla^2 T, \label{eq:th} \\
\nabla \cdot {\bf u} & = & 0, \label{eq:cont_dimension}
\end{eqnarray}
where $\mathbf{u}$ is the velocity field, $T$ is the temperature field,   $P$ is the pressure field, $\nu,\kappa,\alpha$ are the kinematic viscosity, thermal diffusivity, and thermal expansion coefficient of the fluid, and $-g \hat{z}$ is the acceleration due to gravity. 

The first quadrant of Table~\ref{tab:3d_product_rule} describes the product rule for the ${\bf uu}$ and ${\bf u} T$ interactions of RBC.  The rules have been described in Secs.~\ref{sec:Sym} and \ref{sec:magneto_convection}. Here $E$ is the identity element, while the complimentary modes are $M_1' = M_5$, $M_2' = M_6$, $M_3' = M_4$, and vice versa. This is the Klein eight-group. For more details refer to Sec.~\ref{sec:Sym}.  Using the product rules, we can deduce which Fourier modes reverse under a flow reversal, and which ones do not. They are described in Table~\ref{tab:RBC_rules}.

\begin{table} [htp]
\caption{For RBC, classes of Reversing modes (R) and Non-reversing modes (NR).  The remaining modes have small amplitudes and they fluctuate around zero.}
\label{tab:RBC_rules}
\begin{center}
\begin{tabular}{c|c|c|c}
\hline
Item & R  & NR  & classes  \\ \hline 
0 & - & All elements &  No reversal \\
1 & $O$ & $E$ &   7 classes \\
2 & $M_i$ & $E$ &   7 classes \\
3 & $O$ and $M_i$ & $E$ and $M_i'$ &   6 classes \\
4 & $O$, $M_1$, $M_2$ & $E$, $M_3$, $M_5$, $M_6$ &   1 class \\
5 & $O$, $M_2$, $M_3$ & $E$, $M_1$, $M_4$, $M_6$ &   1 class \\
6 & $O$, $M_1$, $M_3$ & $E$, $M_2$, $M_4$, $M_5$ &   1 class \\
7 & $M_1$, $M_2$, $M_5$, $M_6$ & $E$, $O$, $M_3$, $M_4$  &  1 class \\
8 &$M_2$, $M_3$, $M_4$, $M_6$ & $E$, $O$, $M_1$, $M_5$  &  1 class \\
9 & $M_1$, $M_3$, $M_4$, $M_5$ & $E$, $O$, $M_2$, $M_6$  &  1 class \\ \hline
\end{tabular}
\end{center}
\end{table}%

Note that $\{ E, O, M_1, M_5 \}$, $\{ E, O, M_3, M_4 \}$,  $\{ E, O, M_2, M_6 \}$, $\{ E, M_1, M_4, M_6 \}$,  $\{ E, M_2, M_4, M_5 \}$, and $\{ E, M_3, M_5, M_6 \}$ are subgroups of the aforementioned the Klein eight-group.

\bibliographystyle{apsrev4-1}\texttt{}

%

\end{document}